# Mobile Apps for Foot Measurement: A Scoping Review

Muhammad Ashad Kabir[1], PhD;  Sowmen Rahman[3], BS; Mohammad Mainul Islam[4], MSc; Sayed Ahmed[2,5,7], CPed CM Au; Craig Laird[6,7], CPed CM Au;

[1]School of Computing and Mathematics, Charles Sturt University, NSW, Australia
[2]School of Health and Human Sciences, Southern Cross University, Queensland, Australia
[3]Department of Computer Science & Engineering, Chittagong University of Engineering & Technology, Chittagong, Bangladesh
[4]Verizon Media, Sunnyvale, CA, USA
[5]Principal Pedorthist, Foot Balance Technology Pty Ltd, Westmead, NSW, Australia
[6]Principal Pedorthist, Walk Easy Pedorthics, Tamworth, NSW, Australia
[7]Pedorthist, Pedi Wiz Digital Technology, NSW, Australia

**Corresponding Authors:**
Muhammad Ashad Kabir, PhD
School of Computing and Mathematics
Charles Sturt University
Panorama Ave, Bathurst
NSW 2795, Australia
Phone: +61 2 633 86259
Email: akabir@csu.edu.au

## *Abstract*

**Background:** With the public coverage of smartphones now at a global level, a major growth in the use of apps related to the health category, specifically those concerned with foot health can be observed. Although new, these apps are being used practically for scanning feet with an aim to providing accurate information about various properties of the human foot. With the availability of many 'foot scanning and measuring apps' in the commercial market (app stores), the need for an evaluation system for such apps can be deemed necessary as little information regarding the evidence-based quality of these apps is available.

**Objective:** To characterize the assessment of measurement techniques and essential software quality characteristics of mobile foot measuring apps, and determine their effectiveness for potential use as commercial professional tools for foot care health professionals such as pedorthists, podiatrists, orthotists and so on, to assist in measuring foot for custom shoes, and for individuals to enhance the awareness of foot health and hygiene and prevention of foot-related problems.

**Methods:** An electronic search across Android and iOS app stores was conducted between July 2020 and August 2020 for apps related to foot measurement. Mobile apps with stated goals of foot measurement and general foot health were identified and selected by three independent raters and discrepancies regarding the selected apps were resolved via a fourth rater. A modified rating tool based on previous works of app rating tools was adopted and extended for rating of selected apps. The internal consistency of the rating tool was tested with a group of 6 people who rated the selected app over an interval of 5-6 days. This modified scale was then used to produce evaluation scores for the selected range of foot measurement apps and the inter-rater reliability of this study was also calculated. Discrepancies found for any criteria during app reviews were discussed mutually by the raters to arrive at a unified decision.

**Results:** Evaluation inferences found all apps failing to meet even half of the measurement-specific criteria required for the proper manufacturing of custom-made footwear. 23% (6/26) of apps were found to utilize either external scanners or advanced algorithms to reconstruct 3D models of user foot that can possibly be used for ordering custom-made footwear (shoes, insoles/orthoses) and medical casts for fitting irregular foot sizes and shapes. Apps had varying levels of performance and usability, with the overall measurement functionality being subpar with mean 1.97 out of 5. Apps that were linked to online shops and stores (shoe recommending) were assessed to be more usable than other foot measuring apps, but didn't work with custom shoe sizes and shapes. Overall, current apps for foot measurement do not follow any specific guidelines for measurement purposes.

**Conclusions:** Majority of the commercial apps in app stores cannot be comprehensively evaluated as viable apps ready for use as professional tools in assisting foot care health professionals or individuals in measuring their foot for custom-made footwear purposes. Apps lack software quality characteristics that are needed for proper measurement and for providing awareness about foot health and induce motivation to prevent and cure foot-related problems. Guidelines similar to the essential criteria items in this study are needed to be developed for future apps aimed at foot measurement for custom-made or individually fitted footwear and creating awareness about foot health.

**KEYWORDS**
foot measurement app; foot scanning; mobile app; mobile health; custom shoes making; mobile apps survey

# *Introduction*

## Background

Poor foot health is often linked to bad performance both in personal and work life [1]. Various factors that are responsible for the deterioration of foot health originate from anatomical and biomechanical factors, overuse or injury and external trauma. Maintenance of foot health is necessary to keep humans mobile and independent, and consequential negligence can often cause psychological strain along with physical pain [2].

In an aid to combat these problems related to foot health, clinical treatment programs have been widely adopted [33, 34]. These programs consist of clinical interventions, and most of the time require clinicians and patients to have face-to-face contact for over a year. Such interventions have been known to have variable efficacies due to fluctuations in adherences by the patient over time [35, 36, 37]. These rigorous health programs can sometimes be time-consuming, resource and cost intensive, and also can be inconvenient for patients given that foot problems raise a possibility of impeding the movement capabilities of a patient [38, 39]. Accordingly, novel, low-cost and widely accessible tools for accurately scanning and measuring patients' feet and providing health feedback to the patients are needed. This has indeed become a necessity as many patients face significant barriers related to achieving clinical treatments.

The advancement and accessibility of mobile application technology over the recent years has enabled efforts to translate the same traditional clinical treatments and intervention programs towards the development and growth in use of foot health mobile apps. The outcome is the development of mobile apps that can provide insight about patients' feet by leveraging the processing capabilities of mobile sensors such as, multi-sensored cameras, infrared (IR) sensors and features like Augmented Reality (AR). Many such apps utilize algorithms and data-mining techniques to suggest foot and shoe-related solutions based on procured foot measurement values, whereas others stop at providing simpleton information such as, the suggested size of shoe from foot form, and suggested forefoot or toe exercises.

Although, there is an overall increase in app use for feet health conditions, analysis of several apps belonging to this category have led to the discovery about various problems related to usability, design, functionality, lack of free apps, lack of taking user consent and mostly lack of certification and quality of information displayed in order to achieve their most important goals: to improve the patient outcomes [3,4,5]. Consequently, these shortcomings have raised questions about the efficacy and applicability of mobile apps used for foot measurement and scanning [3,4,5].

## Objective

To our knowledge, no studies have extensively explored the current scenario of the commercial mobile app market to review and scientifically evaluate apps related to foot measurement. The abundance and rapid growth of such foot measuring apps in app stores, along with an increased adoption rate of these tools by the common public, demands for an assessment of this rapidly growing market. The objective of the current study was to perform an assessment of published foot measurement apps in the two major commercial app stores (Apple App Store and Google Play Store) by the measurement specific criteria of foot properties and criteria of software quality characteristics and evaluate the viability of these apps for use as professional tools for foot measurement by both pedorthists, podiatrists, orthotists and individual users and investigate their potentiality in causing behavioral changes regarding awareness of foot health and foot related problems.

## *Methods*

### App Search

This study includes the apps found in official mobile app stores: Apple App Store and Google Play Store. An electronic search was conducted between July 2020 and August 2020 in the mentioned app stores for iOS and Android mobile devices respectively. We used an inclusion strategy that consisted of the following: (1) the search process did not take into consideration any subcategories the apps belonged to in the app stores, (2) the terms listed in Table 1 were used to search for foot measurement apps across app stores, (3) apps that weren't indexed on the app stores based on region-based variations were not considered. The methodology used in this study for identification, screening and selection of apps matching criteria is displayed in Figure 1.

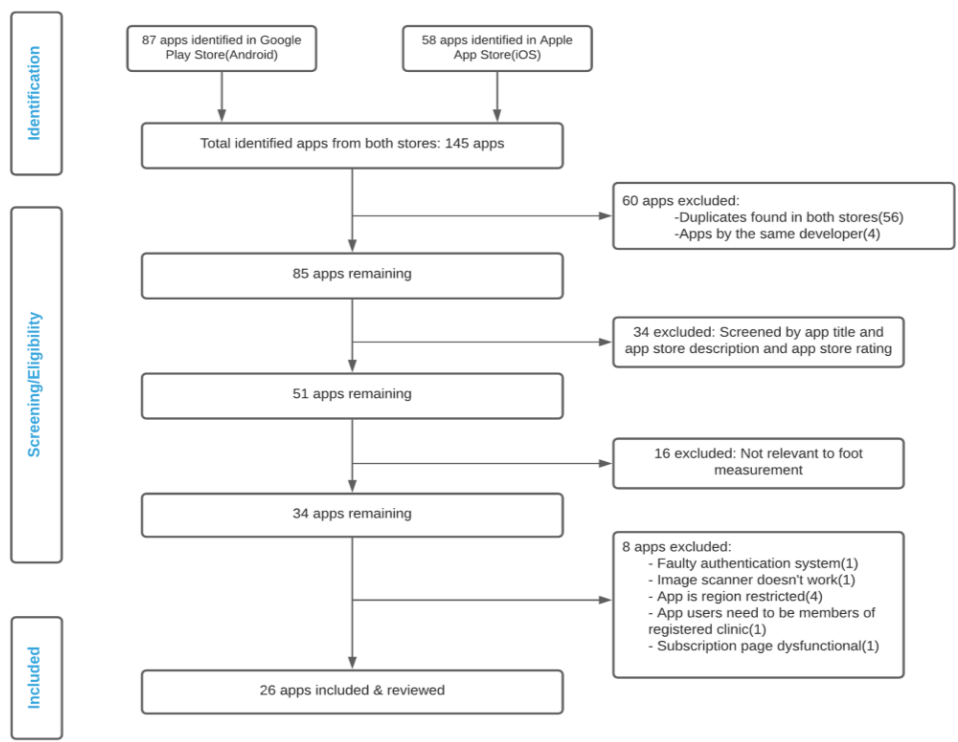

**Figure 1.** Flow diagram of study methods.

**Table 1.** List of terms used for searching app stores.

| "foot scanner", "foot app medical", "foot measure", "feet", "measure foot", "foot length" |
|---|

A secondary search was also executed to identify apps that were intended to be used with advanced imaging sensors and/or other hardware requirements using the keywords listed in Table 2. The results of this latter search were not limited by language, app store description or rating. Both these searches across the separate app stores had no restrictions imposed on them. The search results that were achieved using the enlisted terms shows great variance across app stores. As a result, the exact search using the same terms were executed multiple times across different devices to minimize variance of app indexing and construct the final inclusion list of foot measurement apps.

**Table 2.** Additional terms used for secondary search.

| "foot size", "shoe size", "3d foot", "foot scan solutions" |
|---|

## Selection of Apps

The criteria for selection of feet measuring apps were based on whether the app serves a purpose that involves measurement of the user's foot. The exclusion strategy that was used are pointed out as follows: (1) an app is selected based on whether it supports features to provide information about the individual user's feet or related foot properties, (2) an out of scope requirement includes the usefulness of apps in providing solutions or intervention techniques for preventing or treating major foot diseases, (3) apps that were common to both the app stores i.e. duplicated apps, were verified manually by testing on devices of respective platforms (Android and iOS platforms) and screening off from one store unless that app provided different features that were only supported by a particular operating system, (4) apps falling in appropriate category but had no star rating and no user comments in the store app listing page were not considered, (5) apps were screened by title, store description and store rating relevance as foot measuring apps, (6) app stores were also screened for app duplicates involving the same developer or publisher, (7) app that were inaccessible/unusable due to region restrictions, and (8) apps that had been reskinned with a new user interface over an existing app were considered virtually the same app in terms of functionality and were not considered.

## Modification of existing health app rating tools

We hypothesized that in order to properly rate an app over its appropriateness and usability as a foot measurement app, the app should consist of multiple characterization features to scientifically evaluate the value of the app as a commercially viable foot measurement product. A standardized rating tool will be helpful in this case, especially when there are numerous apps on the app stores. Consequently, an extensive review of prior guidance documents and tools for rating mobile apps was performed to determine the existing fundamental domains and criteria for determination of app usefulness and rating. Having reviewed over a dozen app rating guidance documents and tools, we have come to the conclusion that each one of them has their own uniqueness for particular categories of application. We took into consideration the prior studies on software quality assessment and emphasized on key software quality characteristics such as usability, reliability, functionality, and efficiency [6-9]. We had an aim to further build and extend on the prior models of rating tools such as MARS, uMARS (user-friend version of MARS), and FinMARS (MARS for financial apps), and adapt it to the required evaluation suite for apps used for foot measure [10-12]. With this view in mind, we selected the different categorical domains from aforesaid rating tools, and put forward an extended version of the health mobile app rating system (MARS). This rating tool consists of modifications that we have hypothesized to be of more importance for a foot measuring app and it takes into account individual items that were found during review of the included foot measure apps relevant to the same goal. The finalized rating tool model with the updated overarching domains and individual category items have been illustrated in table 3.

**Table 3.** Extended foot measurement app rating domains and criteria.

| Domain | Criteria |
| --- | --- |
| App Metadata | App platform<br>App store rating<br>App store description<br>App store URL<br>Number of downloads<br>Origin<br>Developer |
| App Classification | App sub-category<br>Applicable age groups<br>App price |
| Aesthetics | Layout consistency and readability<br>Content resolution<br>Visual appeal<br>Group targeting according to app content |
| General app features | Social sharing feature<br>Authentication feature<br>User onboarding interfaces<br>Content customization<br>Visual information<br>Data export options<br>Subscription options |
| Performance and efficiency | Bootup efficiency<br>Accuracy of features and components<br>Responsiveness of app<br>Frequency of app crash |

|  | Overheating device issues<br>Battery life impact |
|---|---|
| Usability | Ease of use<br>Navigational accuracy<br>Gestural design<br>Interactivity & user feedback |
| Measurement specific functionality | Measurement of foot length and width<br>Measurement of foot medial arch height<br>Measurement of foot instep and/or joint girth<br>Measurement of short and/or long heel girth<br>Measurement of heel width<br>Measurement of shoe size<br>Measurement of forefoot tilt/rotation<br>Additional Setup<br>Reconstruction of 3D foot model<br>Additional out of scope features |
| Transparency | User consent<br>Accuracy of store description<br>Credibility/legitimacy of source<br>Feasibility of achieving goals |
| Subjective quality | Overall star rating<br>Overall app purchase preference<br>Overall app recommendation<br>Frequency of use based on relevance |
| Perceived impact of app on users | Awareness induction behavior<br>Knowledge enhancing behavior<br>Scope to improve attitude towards foot health<br>Scope to reduce negligence towards foot<br>Scope to induce foot related help-seeking behaviour |

The key domains that are essential for the evaluation of foot measurement apps have been summarized to the following: app classification, aesthetics, general features, performance and efficiency, usability, application specific functionality, transparency, subjective quality and the app's perceived impacts on users' domains.

The app quality criteria clustered around the domains excluding the metadata section was used to build the app rating scale. Depending on the type of question asked to resolve the criteria, each item can scale a response as a 5-point scale or a binary response (later scaled to a likert scale of 1-5). Cases were found in which it was not possible to acquire information about certain categorical items, as a result of which the "Not applicable" rating was introduced to the scale. Other cases displayed complexity in gaining access to certain types of information, so a rating option 'Unknown' was added. A few subscale items were excluded from quantitative measurement since they provide qualitative descriptions about apps which could not be weighed quantitatively. These items are the app metadata domain items, applicable age group item and app sub-category item.

**App Metadata**
General information about the apps have been abstracted as app metadata from respective app stores under the app classification category. App metadata includes information such as app platform, store URL, store rating, store description, number of downloads, developer information, and origin. These, however, have no impact on the rating scale.

App metadata was extracted systematically by two investigators from each app store on Google Sheets and these datasets were cross-verified for data anomalies.

**App Classification**
Through extensive reviewing of prior works on foot measurement and related technologies about foot measurement [13,14,15,16], the apps were finally sub-categorized depending on the type of functionality into the following: (1) Simple size-unit converter, (2) 2D foot scanner, (3) 3D foot scanner, (4) Shoe recommender, (5) Foot tilt calculator, (6) Foot progress tracker.

Sub-category 1 (simple size-unit converters) apps are the simplest types of app that take in user input values for foot shape and dimensions to produce another category of size or shape, it can be properties related to both shoes and/or foot. 2D scanner apps are more advanced in comparison as they use imaging sensors to acquire 2D data about feet images and use calculation techniques to determine user feet measurements. 3D scanner apps are apps that generally require state-of-the-art techniques and external hardware like 3D imaging sensors and/or dimensional digitizing devices. These types of app generally are capable of providing an array of user feet dimensions. Shoe-recommenders and foot tilt calculators are modified versions of 2D and 3D scanners that do not directly output raw measurement information but transform that into more consumer-friendly, useful views and derived information types. Progress trackers are a common type of apps that do not specifically apply to feet, but they were also included in the study because of their ability to track measurements of different foot areas over time either manually or using in-built measuring techniques.

**Aesthetics**
Today, the market is dominated by hundreds and thousands of apps that are competing over similar categories with similar functions and outcomes but are getting outdone only due to the fact that one app is more appealing to look at, whereas the others are not. Visual appeal is just as important for the success of a commercial product as is the core functionality and performance of that app. Proper layout and organisation of user interfaces elements in an app can sometimes make all the difference between an app's success and downfall in today's competitive marketplace. This trend is also being seen in modern foot scanning apps where apps are being biased based on their visual outlook and how clean and organised the layout of the app is, that is the key element of good interface design is to make it clear and simple for the app users [22].

**General Features**
While providing options for as many measurement dimensions as possible is important for a foot measurement app, there are general features that the app must also account for. Features like the ability to share and export data to other apps and data formats were selected as valid rating items because if it is possible to share data, the user is likely to waste less time on getting their foot size information to other platforms where it may help them. Authentication features are considered a good option to have, when data is stored against a user's credentials, that data is likely to be stored on the cloud, thus removing the user's dependencies on that particular mobile device the app is installed on. Over the recent years, the importance of content customization and the amount of visual information shown in apps has been observed to improve its user value and hence it was included. Additionally, if the app provides subscription packages that may be responsible for affecting the user experience of the app in any way, these should be also factored in.

**Performance and Efficiency**
One of the most important factors that contribute to the functionality of an app is its performance score and how efficiently it can run and provide results on the user's device. The argument is notched up by one point in importance due to the wide scale of customizability of performance components in mobile devices across various global mobile brands. Therefore, as the same foot measurement app may perform differently across different mobile devices with regards to CPU performance, total memory usage, total battery life impact, the possibility of device heating, etc. this domain was included as a criterion for rating foot measurement apps, which are known to use mentionable amounts of processing power and complicated processing algorithms to output the foot dimensions.

**Usability**
Usability of the app refers to the quality of the app's system that is used in achieving the final goals of the app. It is the study of social and behavioral science that is also linked with the science of design [26]. In prior studies involved with human-computer interface (HCI) and user-centered design [27], poor usability and lack of proper user-oriented design has been pointed as 2 of the major reasons why mHealth apps have low user adoption rates. Usability testing of a foot scanning app is very crucial in determining whether the app has a decent enough quality to attract the attention of its target user groups. In today's era of technology, the user's attention in mobile apps is divided between interaction with the mobile app itself and interaction with the environment [24]. Navigation and ease-of-use are important measures of app usability since the screen sequence in-app leads the users through the different views of the app for obtaining the desired information [22]. In prior comparison studies involving comparative app usability testing, it was discovered that the usability of an app in field settings compared to that in laboratory settings varied greatly due to the variance of user behavior and the user experience [24]. This signifies that the usability testing of an app is a necessary stage of the app development cycle. Taking view into prior works of app rating systems [10,11,12], we hypothesized to assess an app's usability as good given that: (1) the app has the ability to be operated with ease, (2) the app's navigation flow is uninterrupted, (3) the gestural design (if present in app) and screen links(buttons, arrows, navigation panels, etc.) is consistent across all app pages, (4) the app provides an interactive experience by taking input from the users and giving feedback when necessary.

**Measurement specific functionality**
In a foot measurement specific app, the dimensionality of features provided by the apps are very important. In simple terms, if an app A can measure more foot properties than another app B, then the potential utility of app A can be considered more than that of B. Strategizing on this, we have decided to include different types of apps that are directly or indirectly involved with the process of foot measurement in this study. We have reviewed many works on foot dimensions' measurement. Currently, foot dimension extraction has become heavily dependent on 3D scanners. These types of scanners are currently being used in commercial and research areas, some specifically for the measurement of foot dimensions [13]. Other studies have explored various measurement techniques using digital light project technology, image sensors, and 3D digitizing devices including second generation Kinect [14,15,16]. In their techniques, they measured foot length, foot width, metatarsal/ball girth. Again, for footwear and insole design, instep and medial arch height, ball girth and forefoot tilt are also necessary [17,18,23]. In various other techniques, laser scanners are used for scanning foot length. This scanned data can also be refined and modified using the laser scanners and used for re-modelling of the human foot [19,20,21].

Therefore, taking these important pedorthic guidelines for the measurement of the foot into account, the app measurement specific functionality category considered for the weighing of foot measurement apps was composed of the following measurement properties: (1) foot length, (2) foot width, (3) arch height, (4) instep girth, (5) joint girth, (6) short heel girth, (7) long heel girth, (8) heel width, (9) shoe size, (10) forefoot tilt. Additional discoveries about taking input from camera sensors/images, requirement of calibration markers or extra setup, along with the possibility of reconstruction of the 3D model of foot were scoped into the rating scheme as functionality subscale items.

**Transparency**
It is a known fact that mobile apps that utilize social and personal private information for their proper functioning, are common targets for various businesses that capitalize on personalised services [25]. Often, apps sell private information critical to the normal livelihood of individuals without awareness and the main cause of this is due to improper mobile privacy decision making. It is to be made sure that when a user gives their consent to private data being accessed by the apps, that even though users may or may not know the direct consequences of such actions, the apps themselves strictly follow specific forms of data protection and regulation rules and explicitly express to their users how and why their data is being collected. In prior studies [25], tests regarding apps' purpose and terms of private data processing use on users have been conducted and positive results have been achieved thus suggesting that expressing user consent and following data protection rules are viable items of transparency domain of an app. In case of foot measuring apps, the aforesaid constraints should be followed along with verification of the publisher or developer, whether the source of the app can be trusted and whether the app is successful in meeting its goals as described in the store description, which is also subject to scrutiny. With such information, a user can make an informed decision beforehand about downloading the app by determining the authenticity of the app.

**Subjective Quality**
App subjective quality refers to the individual app user's key perspective views about the app. These can be anything between personal app ratings, good and bad comments about the app, preference to pay for an app based on its features, preference to recommend an app and use an app based on relevance to the user. Often, a general direction about what the app offers on the app store can be guessed by just seeing how the app store users who have previously downloaded and used the app have reacted about it. This is however subjective, since the general distributed value of app comments and ratings saturate towards an approximated value only as the number of the reviews become large for an app, thus, this approach to measuring the performance of an app pre-download is not applicable for apps with few or no user ratings and/or comments on its app store. However nowadays, users are often observed to comment about apps comparatively in-depth with key points that are helpful during the app review phase, as a result of which this is an optional but valid criteria item for apps retaining that saturated direction from user reviews and comments.

**Perceived impact of app on users**
When an app is consumed by its users, the impact of that app on its users becomes a notable indication of how useful that app is potentially. Development in technology and continued breakthroughs in the branches of computer science and machineries have helped thousands of users in guiding their own health to safety and prevented deaths, with a large number of apps designed to improve user health [28]. In mHealth apps, the app's main scope to induce awareness about a particular health problem and increase motivation to avoid and prevent future occurrences related to health are regarded as one of the main objectives. Additionally, mHealth apps may also provide intervention techniques and advice useful in decreasing the user's negligence towards health and increase help-seeking behaviors targeting solutions to health problems. Pertaining to the same category(health) of mobile apps, the same views were held for foot measuring apps. However, Milne-Ives et al. [29] have found that most of the apps that are user health-oriented yield little to no evidence of effectiveness in cases of patient health outcomes and health behavioral changes. To conclusively support the foot measurement apps as useful tools for changing outcomes of foot health and attention-based behavior, these apps must be evaluated by their effectiveness value on the app users [29].

## Internal Consistency of Modified Scale and Inter-Rater Reliability
We used Cronbach's alpha [31] for calculating the internal consistencies of the overarching domains of the modified rating scale: aesthetics, general app features, performance and efficiency, usability, measurement specific functionality, transparency, subjective quality and perceived impacts on users. For this work, one of the foot-measurement apps "INESCOP YourFeet" was selected for review of this study, was additionally used to rate the internal consistency of our modified rating scale. For all independently participating raters, the internal consistency of all subscale items was good to high. The overall internal consistency of the modified rating scale was high at alpha = .84 which is considered excellent according to prior studies [32] The subscale alpha's were also in the range of good-excellent (aesthetics alpha = .88; general features alpha = .92; performance alpha = .7; usability alpha = .84; functionality alpha = .92; subjective alpha = .79; transparency alpha = .9; perceived impacts alpha = .78).

For the measurement of the inter-rater reliability of our raters who reviewed the set of 26 apps in this study independently, we used the previously mentioned method in an analogous manner. The inter-rater agreement score is ranged between .54 to .70 and is considered as a *fair-good* level of rater reliability or agreement [32].

# Results

### Summary of Search Results
A systematic approach was used to search for apps in the app stores. The initial searches using the terms from table 1 and table 2 yielded a result set of 145 apps across the two app stores, out of which, 41.4% (60/145) of apps were excluded from any further review for satisfying the exclusion criteria of app duplication across stores. Of the remaining apps. 23.4% (34/145) of the apps were excluded using the exclusion criteria of irrelevant title, store description and store rating. From the rest, after installing and using the apps, 11.0% (16/145) of these apps failed to satisfy the foot measurement category requirement. Out of the remaining 34 apps, 8 apps were excluded for having faulty authentication systems, subscription pages, region-restrictions, not being registered clinic members and technical issues in image scanning that were crucial for access to the app features. In total, 26 apps were selected as eligible to be reviewed as foot measuring apps by our proposed modified rating scale.

### Overall Assessment of Apps
The overall assessment of all eligible apps in this study listed in table 4 has led to the discovery of most apps belonging to the 2D and 3D scanning sub-categories. The categorical distribution of all reviewed apps can be seen in Figure 2.

These two categories are the most important ones since they handle output as raw measurement values which may be used for making custom-made shoes or recommend individualised shoe fit, and providing users with measures of foot dimensions regarding their feet that may further be used for maintaining foot health and preventing foot related problems. Apps of these two categories were dependent on calibration markers such as standard-sized (A4, A5, etc.) papers and purchasable barcode stickers and utilized the camera sensors on mobile devices to capture, process and measure the foot dimensions of users. 30.8% (8/26) of the apps were of 2D scanning type, and 34.6% (9/26) were of 3D scanning type. Seven of 10 3D scanning apps required an external sensor (Structure sensor, KinectV2, etc.) for proper functioning and all of these apps could reconstruct the 3D foot model which is a good point since data about the 3D foot model is reusable for making custom shoes.

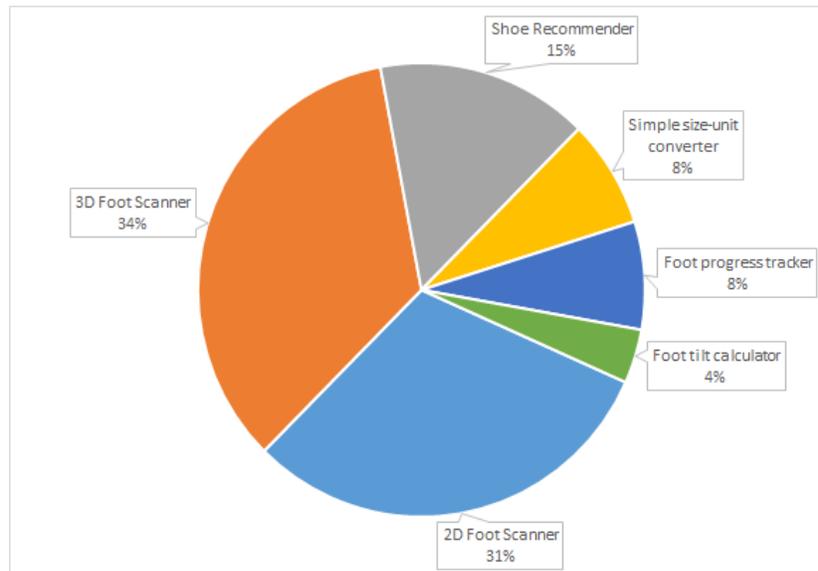

**Figure 2.** Categorical distribution of reviewed foot measurement apps

2 out of 26 apps were of the simple unit converter category and hence their general features score, and functionality scores were not calculated as the processes performed by these apps did not involve directly taking input of user foot measurements via scanning or answering questionnaires. Majority of the apps targeted a general group of users with no age requirements (17/26, 65.4%), while apps specifically categorised as 2D and 3D foot scanning apps (7/26, 26.9%) targeted young adults and adolescents. There were 2 apps which dealt with users of younger age groups (aged between 2 to18) and these are the apps "Jenzy: Easy Kid Shoe Sizing" and "Remeasure Body". The app "Remeasure Body" was however not a 2D or 3D scanning app, it was a foot progress tracking app. Assessment revealed many commercial apps that used advanced image scanning and processing techniques for determination of foot size, shoe type and size and even reconstructed 3D models of foot for getting the size of shoes and insoles (ATLAS - Scan your feet!, ECLO, Fischer Scan-Fit, Nimco Professional Shoe Sizing, Jenzy: Easy Kid Shoe Sizing, ShapeCrunch).

From calculated results, it is seen that the overall navigability, design and appeal of current apps score better compared to other aspects of the app, with the general range of usability standing between 3.75 to 4.50. But the general range of measurement specific functionality was low for all apps except the app "Nimco Professional Shoe Sizing". This app performed better with the abilities of measuring both foot length and width, foot instep height, and ball girth. The app also included the 3D model reconstruction of foot using the procured measurement values. An additional feature of the app is that it can suggest shoe shape and size based on the processed 3D foot model.
Except one app, the rest in the list were free to download, with 30.7% (8/26) of apps having subscription packages available which were essential for getting full access to the apps' suite of functionality (limited free).

**Table 4.** Assessment scores for foot measurement apps

| Name | Aesthetics | General | Performance | Usability | Functionality | Subjective | Transparency | Impact | Total Mean |
|---|---|---|---|---|---|---|---|---|---|
| ShapeCrunch | 4.00 | 3.29 | 4.33 | 4.00 | 2.33 | 3.75 | 4.67 | 4.40 | 3.85 |
| INESCOP YourFeet | 5.00 | 3.00 | 4.67 | 4.75 | 1.73 | 3.25 | 5.00 | 2.60 | 3.75 |
| FISCHER Scan-Fit | 5.00 | 3.00 | 4.17 | 4.00 | 2.45 | 3.75 | 4.67 | 3.00 | 3.75 |
| Foot Measure | 2.50 | 1.50 | 3.83 | 3.75 | 1.73 | 1.00 | 1.33 | 1.00 | 2.08 |
| SizeMyShoe | 4.25 | 3.00 | 4.67 | 4.00 | 1.36 | 2.25 | 3.67 | 2.40 | 3.20 |
| ATLAS - scan your feet! | 4.25 | 3.00 | 3.33 | 4.25 | 1.36 | 4.00 | 5.00 | 2.20 | 3.42 |
| Jenzy: Easy Kid Shoe Sizing | 5.00 | 3.29 | 4.67 | 5.00 | 1.36 | 4.25 | 5.00 | 2.00 | 3.82 |
| Shoe Size Meter - foot length | 4.25 | 3.00 | 4.67 | 3.75 | 1.36 | 3.50 | 4.33 | 3.60 | 3.56 |
| Shoe Size Converter | 4.50 | - | 5.00 | 4.50 | - | 3.50 | 4.67 | 2.80 | 3.80 |
| The Foot Fit Calculator(BikeFit) | 5.00 | 2.14 | 4.50 | 4.50 | 1.67 | 3.00 | 4.00 | 3.60 | 3.55 |
| Foot Length Converter Size in Lite | 4.00 | - | 5.00 | 4.00 | - | 2.50 | 4.33 | 2.20 | 3.39 |
| myFoot - Rescue your feet! | 4.25 | 4 | 4.67 | 4.75 | 3.00 | 4.00 | 4.75 | 4.40 | 4.23 |
| Remeasure Men Body | 4.75 | 4.00 | 5.00 | 4.50 | - | 4.75 | 5.00 | 3.20 | 4.07 |
| FootFact | 3.25 | 2.00 | 4.67 | 4.25 | 1.67 | 1.50 | 2.33 | 1.60 | 2.66 |
| Swift Orthotics | 4.50 | 1.50 | 4.00 | 4.25 | 2.09 | 4.00 | 3.00 | 3.20 | 3.32 |
| Nimco Professional Shoe Sizing | 5.00 | 4.20 | 4.50 | 4.50 | 4.33 | 4.50 | 4.67 | 3.40 | 4.39 |
| Shoe-buddy | 4.25 | 3.67 | 4.17 | 4.00 | 2.33 | 4.00 | 4.75 | 3.60 | 3.85 |
| 3D Avatar Feet | 4.25 | 2.14 | 3.67 | 4.25 | 2.67 | 4.25 | 4.75 | 3.20 | 3.65 |
| SUNfeet | 3.75 | 2.71 | 4.00 | 3.75 | 2.67 | 3.75 | 5.00 | 2.80 | 3.55 |
| ECLO | 5.00 | 3.86 | 4.33 | 4.75 | 2.33 | 4.00 | 5.00 | 2.40 | 3.96 |
| FotAppenScan | 1.50 | 1.57 | 4.50 | 2.50 | 1.67 | 2.25 | 3.00 | 2.20 | 2.40 |
| Ortholutions | 4.75 | 1.50 | 4.33 | 4.50 | 1.67 | 4.25 | 5.00 | 4.20 | 3.78 |
| AARA Orthotics | 4.50 | 3.50 | 4.00 | 4.25 | 1.67 | 4.00 | 4.75 | 4.40 | 3.88 |
| Aqualeg | 1.75 | 2.00 | 3.67 | 1.50 | 1.67 | 2.25 | 2.00 | 1.20 | 2.00 |
| Anodyne Scanner | 4.50 | 2.00 | 4.00 | 4.50 | 1.67 | 4.25 | 4.75 | 3.40 | 3.63 |
| 3DsizeMe | 4.75 | 3.00 | 4.33 | 5.00 | 1.67 | 4.50 | 4.75 | 3.40 | 3.93 |

The mean overall app rating was 3.52 out of 5 (95% CI, 3.32-3.71) (Fig. 3). Among the domains, significant differences were detected, most notably with impact, general features and functionality receiving the lowest rating by far: 2.94, 2.79 and 1.97 out of 5. In contrast, the most highly rated domains were performance and transparency receiving 4.33 and 4.24 out of 5. The error bars represent 95% confidence intervals for each domain. Other domains that scored with higher mean values are aesthetics (4.17/5) and usability (4.14/5).

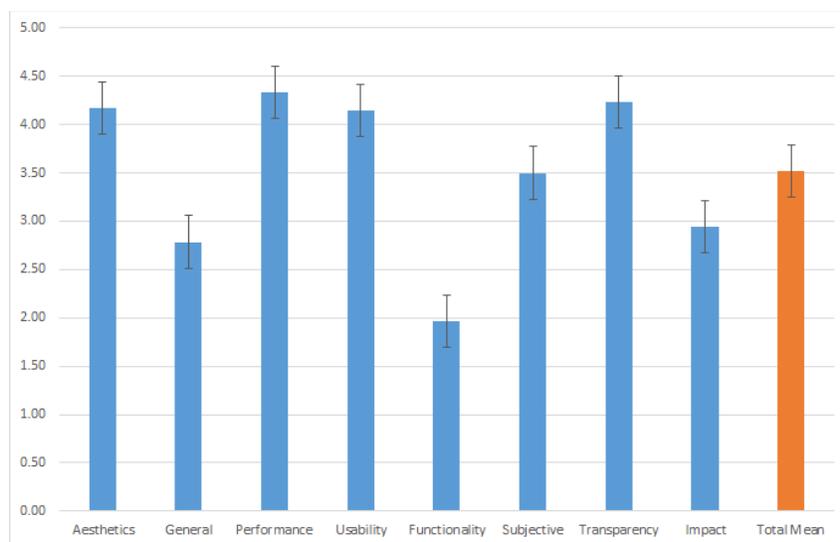

**Figure 3.** Overall app ratings

## Comparison of store rating and rating scale measured ratings

The store ratings of reviewed apps were compared to the score of the apps from our rating scale (illustrated in figure 4). The standard deviation of the difference of two scores for reviewed apps was 1.07. This deviation is not too poor considering that the score in our rating scale is an aggregated mean of various domains that are necessary for specifying the quality and criteria of foot measuring apps. Even though the ratings were within a close range of spread, there is need for more than just a store rating to drive users to installing and using the apps since the category of apps that are part of this study aren't intended to be used as casual apps(for the evaluation of the study's objective). During the review, no star ratings for apps SUNFeet, Ortholutions, AARA Orthotics, and Aqualeg were found and thus were excluded from the calculation of the total std. deviation.

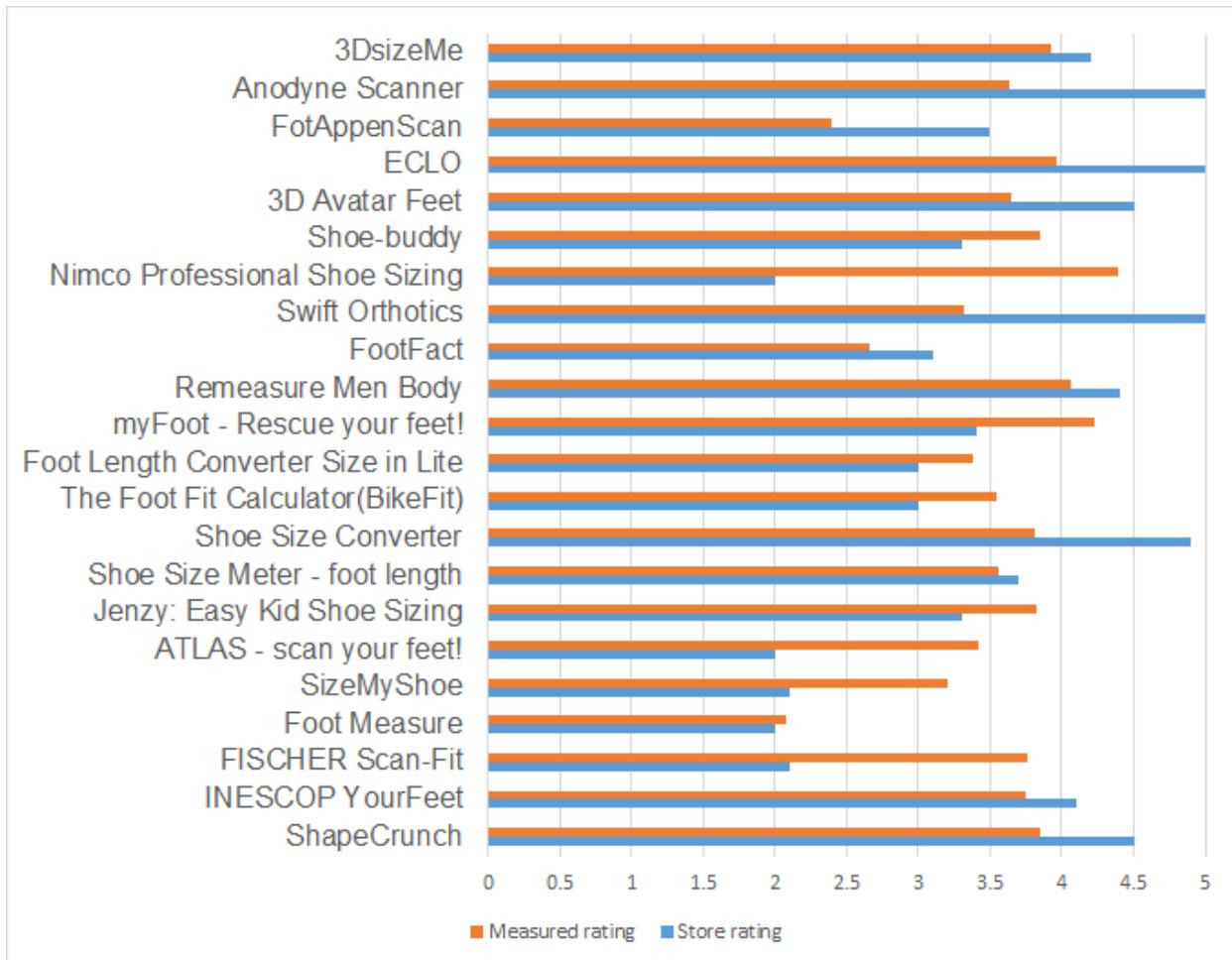

**Figure 4.** Comparison chart of app ratings from app store and developed rating scale

## Assessment of Measurement Criteria in Apps

For foot measuring apps, the keynote function is the measurement of foot dimensions that keep the users aware about the current condition of their feet. In accordance with the selected measurement criteria for foot measurement apps, an evaluation of the comprehensiveness of reviewed apps regarding the criteria was performed, portrayed below in table 5. The assessment showed that there was a great deal of variation in the number of foot dimensions measured by the apps. Table 5 illustrates a transformed view of the quantity of measurement items of apps as discussed in the measurement specific functionality subsection of the methods section. During calculation of the results of this table, apps that were identified as simple size-unit converters and did not utilize the feature of taking images of the user's foot for scanning and measuring purposes, we used the value of "Not applicable" for various measurement subscale items related to that particular feature so that there existed no calculation inconsistencies.

**Table 5.** Assessment criteria for measurement functionality of apps

| Measurement Criteria | Google Play Store (n=11) n(%) | Apple App Store (n=15) n(%) | Total (n=26) n(%) |
|---|---|---|---|
| Foot length | 6(54.55%) | 7(46.67%) | 13(50.00%) |
| Foot width | 6(54.55%) | 5(33.33%) | 11(42.31%) |
| Foot arch (medial) height | 1(9.09%) | 2(13.33%) | 3(11.54%) |
| Foot instep girth | 0(0.00%) | 1(6.67%) | 1(3.85%) |
| Foot joint girth (ball girth) | 0(0.00%) | 3(20.00%) | 3(11.54%) |
| Short heel girth | 0(0.00%) | 0(0.00%) | 0(0.00%) |
| Long heel girth | 0(0.00%) | 0(0.00%) | 0(0.00%) |
| Heel width | 0(0.00%) | 2(13.33%) | 2(7.69%) |
| Shoe size | 2(18.18%) | 2(13.33%) | 4(15.38%) |
| Forefoot tilt | 0(0.00%) | 2(13.33%) | 2(7.69%) |
| 3D foot model reconstruction | 1(9.09%) | 10(66.67%) | 11(42.31%) |

From the result in table 5, it is observed that the general distribution of measured foot dimensions in current apps is very low. From the total of 26 apps, only 50% of the apps measure foot length. The percentage values for the remaining dimensions are: 42.3% apps measure foot width, 15.4% apps can determine shoe sizes for users, 11.5% apps can measure foot instep height and ball girth, 3.85% apps are able to measure foot instep girth and heel width. Important dimensions that are crucial for the construction of shoes, insoles and foot wedges (for bicycle pedals, saddles, etc.) were missing from the measurement features offered by all reviewed apps. Most of the apps that were reviewed from the Google Play Store were found to be of comparatively poorer quality than those collected from the Apple App Store. However, since apps were excluded on the basis of duplication across app stores, and region-restriction criteria for this study, this information should not be used to reflect the current state of platform-specific apps, i. e., Android devices have a lesser variety of apps devised for foot measurement compared to iOS devices.

Further evaluation of the measurement criteria showed that the maximum number of dimensions that were measured in a foot measuring app was equal to 5. No app measured more than half of the required foot measurement criteria and a majority (50%, 13/26) of the apps only measured 1 dimension. There were 3 (11.5%) apps that measured 5 dimensions, 3 (11.5%) apps measured 4 dimensions, 1 (3.85%) app measured 3 dimensions, 3 (11.5%) apps measured 2 dimensions and 3 (11.5%) apps that measured no dimensions at all. In general, most of the reviewed apps were not suitable for foot measurement in clinical practice for custom-made shoes and insoles/orthoses or for general use for individuals.

## Analysis of User Reviews from App Store

It is not quite recent that app stores have introduced the ability to let users review apps hosted by them. These user reviews provide a rich source of information about the performance and future viability of apps. Hence, to drive the market competition of apps (for any category) up the ladder of commercial success, developers and publishers aim to receive good and positive reviews as these are the crowdsourced quality indicator of apps [30].

For this study, user comments from the app listing page from respective app stores were collected along with the app metadata for analysis. The user comments were divided into two types based on the user's review rating of the app: the comment is GOOD if rating is 4 stars or above, otherwise BAD. Analysis of the user comments regarding foot measurement apps from both the stores leads to the discovery of most of the apps lacking good feedback or reviews from app users since the number of the downloads in related apps were generally low, and an estimated 30.8% (8/26) of apps had an audience of 100 to 1000 users. In the case of iOS apps, the number of downloads could not be viewed. However, in this case too, the number of reviews with information by users were too low. These findings are conclusive with the findings of studies by Vasa et al. [30] that users tend to write little to nil lengths of information for the reason of their rating when an app performs better in its category. In our case of review, we found an approximate 42.3% (11/26) percent of apps to have very short or no informative description along supporting their ratings.

46.1% (12/26) of reviewed apps contrasted otherwise with detailed reviews from users. One of the most positive sets of user responses was found from the app "ShapeCrunch". This app is a foot scanner that uses machine-learning generated 3D printed insoles to reduce the discomfort of foot pain. Based on the reviews of several users, this app is rated "excellent" on the ground that it can provide properly fitted insoles for users with a wide variety of foot problems. Another app named "Swift Orthotics" also scored a good rating that was consistent with overall ratings both in the app store and in modified evaluation scale. This app used AR technology to measure foot dimensions, and was the only other app that could measure heel width along with FISCHER Scan-Fit among all other reviewed apps. Strong positive feedback with no drawbacks were received from apps ECLO and 3DSizeMe, which were positively correlated with the ratings of the independent raters, although ECLO was a shoe recommending app that can order shoes based on user scans from the ECLO online store. The app "3DSizeME" has great support for Structure Sensor, an advanced 3D scanner, and already there are many companies who directly support 3DSizeMe file formats for ordering custom shoes and insoles. On the other hand, apps such as "Aqualeg", "Ortholutions", "AARA Orthotics" which used external scanning mechanisms didn't receive much public exposure although they scored good ratings according to the devised rating scale, and all these apps had features for visualizing, and either uploading foot scans to company servers or exporting the model data to order shoes, insoles and even medical casts for different body parts. Similar results were found for the 2D scanning app "INESCOP YourFeet" which didn't have good store descriptions but store ratings were consistent with our measured ratings, this app however lacked good functionality of providing detailed foot information for making custom footwear(although it can measure regular shoe size).

Inconsistencies were found among app user reviews with the actual performance of the apps by our raters. These apps are "Shoe Size Meter - foot length", "SizeMyShoe", "FISCHER Scan-Fit", "Jenzy: Easy Kid Shoe Sizing", "INESCOP YourFeet", "The FootFit Calculator". Two of the apps that performed better than the rest 3D scanner apps (as reviewed) were "SUNFeet" and "3D Avatar Feet", however, their respective app store pages had no user reviews at all.

From the above analysis of foot measurement app user reviews, we conclude about the situation of ratings and reviews of current apps by pointing at the inconsistency of rater app ratings (in this study) with app store user reviews. This may be possible due to the app's performance and functionality being dependent on the specific device version and software being used. This suggests that a majority of the current published apps suffer from device architecture and build-related problems and need to be optimized. Another possible cause of this rating versus review inconsistency is that users while rating apps on the store do not generally focus on domains like perceived impacts, transparency, and technical functionality of the apps thus making the point of commenting about the app's features and lacking even more difficult to reach.

## Discussion

### Principal Findings
The findings of this review can be segmented into three major sections: (1) the viability of apps in podiatric practices for making custom shoes, (2) the viability of apps for individual use for general measurement purposes, (3) the potential of inducing behavioral changes about foot health and foot related problems among users.

This review demonstrates that although there are a handful of foot measurement apps available in the commercial stores, the performance of apps with regards to the objective(s) of this study are found to be poor and with lacking features. The objective of current apps being used in clinical practice for custom-made shoes is far from being achieved with the current configuration of foot measuring apps available in app stores. While apps may be used for different types of uses such as, online shopping for shoes and insoles, and casual measurement of feet, they are deemed unusable as pedorthic tools for measurement of foot dimensions which require a comprehensive fulfillment of measurement criteria determined by this study for proper and precise measurement for custom-made shoes and insoles.

Most reviewed apps did not contain half of the required measurement dimensions criteria for properly measuring the feet of users. Some of the apps that performed measurement of foot in the app didn't provide enough relevant information about the actual foot dimensions they measured, but rather this information was used and processed for other purposes. No app included any information about the degree of accuracy that can be achieved from the measurement using the technologies that were used.

In general, most of the current apps belonging to 3D categories could in fact reconstruct the 3D model of feet. Some apps would measure a small number of foot properties and send scanned information to their internal servers for 3D reconstruction of foot model, some would take in information and use those to find and fit custom-made shoes and also insoles/orthoses for users (possibly with foot-related disabilities and zonal pain), and have them delivered at the doorstep of users. But, except 6 (23%) apps (Nimco Professional Shoe Sizing, Fischer Scan-Fit, 3D Avatar Feet, SUNFeet, 3DSizeME, Anodyne Scanner), no other app reported the user's foot dimension information with enough detail which can be used for making custom-made shoes (and insoles). The app 3DSizeME is mentionable due to its performance and data shareability options, which are lacking in the majority of current apps thus making the mobility of data difficult. In general, apps tended to focus on particular measurement criteria which cannot completely describe the structure of foot for making custom-made shoes.

For the general measurement use of individuals, this study finds a majority of the reviewed apps being either developed for commercial purposes and most likely to be used as lookup apps for getting preferred foot/shoe sizes. However, the usage frequency of apps underperforms by a large margin. 65.40% (17/26) of the apps subjectively reviewed by raters had an average subjective usage frequency of under 10 times over the period of 12 months. 30.80% (8/26) of the apps had usage frequency of 10 to 50 and only 1 app was properly usable with usage frequency of 50+ over 12 months. Apps of the shopping category that met this usage range are: ATLAS - Scan your feet!, Shoe Size Converter, and Nimco Professional Shoe Sizing. The usage statistics and calculated ratings of such apps in this review is conclusive with the analysis of the user reviews from app stores as well, suggesting that the cause of poor usage of apps, even though a large percentage of these apps related to foot measurement were available free of charge and mostly provided all included features freely, can be a result of lack of more features and presence of poorly optimized features in the apps. Thus, for individual use too, the apps are not likely to be used as foot measuring tools of high value. Some apps had detailed complaints about their inability to properly size feet and shoes, while others suffered from performance issues including battery draining, network errors, and overheating problems over prolonged use of apps. In some apps, structural flow of technical aspects was confusing and difficult to follow through for the users.

An inference of this review demonstrates that the apps that derived shopping category related results from the calculated foot measurements performed better overall in comparison to foot scanning apps that outputted raw measurement values. There was a total of 7out of 26 apps that met this criterion and all apps were measured to have consistently good mean scores.

Based on the accuracy of their store description and credibility of the app developers, apps were mostly consistent about their intended use, however some apps were not explicit about consent of use of data. Some apps displayed warning dialogs within apps to notify users about providing private data access, whereas some apps didn't even have enough information about their protection policy of personal private information on their support websites or inside the app privacy policy page (if relevant), which raises a question about the authenticity of their intention of keeping user data private.

The overall perceived impacts about current apps shows that apps have low fidelity for bringing behavioral changes when it comes to promoting foot health and awareness of foot related problems. The apps mostly didn't meet a majority of the required basic measurement dimensions needed for measuring feet properly and hence were deemed as not suitable for inducing health-related awareness (79.3% app responses were negative) and help-seeking behavior (82% app responses were negative) related to foot health and problems. However, the study showed that most of the apps that were used in cases other than direct measurement of foot dimensions did have a positive effect (64.7% positive responses) on the necessity of buying properly sized shoes and insoles. Such impacts suggest that while the commercial market has grown with the increase of foot measurement related apps, technical quality of the apps need more improvement and developers should carefully follow guidelines like the ones provided by our rating scheme while developing apps publicly available for suiting custom-made footwear or individualised shoe fit for all types of feet.

## Limitations
The search methods that were used in this study follow a modelling pattern similar to previous studies that were involved with various mHealth categories including management of various mental and chronic diseases, management of food diet, use of drugs, alcohol, physical activity, weight, and management of diabetes [3,4,5,27,29]. This work focuses on apps which were not access-restricted by region. The comprehensiveness status of apps related to various domains of software characteristics that are hypothesized to be of more importance for foot measurement were presented through this study and the devised rating scale was created targeting the practice of thorough review of published foot measurement apps. Although these apps were not verified of presence of applied knowledge of foot morphology and foot health, a general impact value of the apps were taken to preview how the current consumer party is welcoming the features provided by these and whether these features are capable of producing a sense of awareness about foot health. It should be noted that in spite of testing all apps selected and eligible for the review by raters independently, these values should not be interpreted to focus on any particular criteria or item, as this was not the main objective of this study. The findings presented are a broad characterization of the general quality characteristics and measurement-specific features that should be present in foot measurement apps for creation of custom footwear products and general use by individuals for providing awareness of foot health, and are meant to represent the current state of the commercial app market with regards to foot measurement apps. The current study is meant to provide a view about the knowledge of developers and publishers regarding technical knowledge about foot health and foot measurement, and point at the possible directions of advancement of research and development in the foot health and measurement categories.

## Future Directions

The analysis made for this study is not completely perfect because although we tried to assess all important software characters of apps that were important from different points of view, certain statistical information about reviewed apps such as, the accuracy of measured dimensions of feet, CPU, memory and battery usage of the app could not be assessed due to resource constraints and future research will focus on determining the accuracy achieved by reviewed apps and alongside introduce more app stores for an even more comprehensive review. One of the future recommendations for this study would also be to investigate the possibility of including more features as part of the categorical criteria that would ensure more robust foot measuring apps with enhanced technical features. The tool that has been developed by extending from existing high-quality rating tools may be useful for both developers to address the major issues found in this review: measurement-specific functionality, transparency, performance and stability. Another direction to this study would be a direct extension and update to the review since, among the vast collection of health apps we have reviewed for this study, we may have missed apps due to our search criteria, and even there may be apps that were not available due to regional restriction of app stores.

## Conclusion

The verdict drawn from this review about foot measurement apps is that in its current state, the majority of mobile apps in app stores do not meet sufficiently with the criteria required for manufacturing custom-made footwear or to recommend individualised shoe fit. While a few apps do indeed provide enough information about user feet that may be manually acquired from the apps for the objectives of this study, we believe that only by addressing the entirety of issues found and applying more caution to implementing features that are required for the completeness of foot measurement, developers will be able to bring useful apps to the commercial stores that will be eligible for direct professional use in clinical practice for custom-made footwear, alongside motivate individuals to properly address foot problems and become more aware of the importance of foot health.